\documentclass[10pt,letterpaper]{article}
\usepackage[top=0.85in,left=2.75in,footskip=0.75in]{geometry}

\usepackage{amsmath,amssymb}

\usepackage{changepage}

\usepackage[utf8x]{inputenc}

\usepackage{textcomp,marvosym}

\usepackage{cite}

\usepackage{nameref,hyperref}

\usepackage[right]{lineno}

\usepackage{microtype}
\DisableLigatures[f]{encoding = *, family = * }

\usepackage[table]{xcolor}

\usepackage{array}

\newcolumntype{+}{!{\vrule width 2pt}}

\newlength\savedwidth

\newcommand{\beginsupplement}{%
        \setcounter{table}{0}
        \renewcommand{\thetable}{S\arabic{table}}%
        \setcounter{figure}{0}
        \renewcommand{\thefigure}{S\arabic{figure}}%
     }

\raggedright
\usepackage[aboveskip=1pt,labelfont=bf,labelsep=period,justification=raggedright,singlelinecheck=off]{caption}

\makeatletter
\renewcommand{\@biblabel}[1]{\quad#1.}
\makeatother

\usepackage{lastpage,fancyhdr,graphicx}
\usepackage{epstopdf}

\usepackage{xspace}

\fancyheadoffset[L]{2.25in}
\fancyfootoffset[L]{2.25in}
\newcommand{\dd}{$\delta$D\xspace}
\newcommand{\dox}{$\delta^{18}$O\xspace}

\newcommand{\kybp}{ka\xspace}

 \usepackage{wasysym}

\begin{document}
\vspace*{0.2in}

\begin{flushleft}
{\Large
\textbf\newline{Climate entropy production recorded in a deep Antarctic ice core} 
}
\newline
\\
Joshua Garland\textsuperscript{1\Yinyang *},
Tyler R. Jones\textsuperscript{2\Yinyang},
Elizabeth Bradley\textsuperscript{1,3},
Michael Neuder\textsuperscript{3},
James W. C. White\textsuperscript{2},
\\
\bigskip
\textbf{1} Santa Fe Institute, Santa Fe, NM, USA
\\
\textbf{2} Institute of Arctic and Alpine Research, University of Colorado at Boulder, Boulder, CO, USA
\\
\textbf{3} Department of Computer Science, University of Colorado at Boulder, Boulder, CO
\\
\bigskip

%
%
\Yinyang These authors contributed equally to this work.





* joshua@santafe.edu

\end{flushleft}
\section*{Abstract}
Paleoclimate records are extremely rich sources of information about
the past history of the Earth system.  Information theory---the branch
of mathematics that addresses the task of quantifying the degree to which the present
is informed by the past---provides a new means for studying these
records.  We demonstrate that estimates of the Shannon entropy rate of
water-isotope data from the West Antarctica Ice Sheet (WAIS)
Divide ice core, calculated using weighted permutation entropy (WPE),
can bring out valuable new information from this
record~\cite{joshua-ida-2016}.  We find that WPE correlates with
accumulation, reveals possible signatures of geothermal heating at the
base of the core, and clearly brings out laboratory and
data-processing effects that are difficult to see in the raw data.
We also find that  signatures of Dansgaard-Oeschger events in the
information record are small, suggesting that these abrupt warming
events may not represent significant changes in the climate system
dynamics. While the potential power of information theory in paleoclimatology
problems is significant, the associated methods require careful
handling and well-dated, high-resolution data.  The WAIS Divide ice
core is the first such record that can support this kind of analysis.
As more high-resolution records become available, information theory
can become a powerful forensic tool in climate science.


\section*{Introduction}

The Earth contains a vast archive of geochemical information about the
past and present states of the climate system.  The data in these
records---samples from corals, marine and lake sediments, tree rings,
cave formations, the ice sheets, etc.---captures a rich spatiotemporal
picture of this complex system.  Ice cores, for example, provide
high-resolution proxies for hydrologic cycle variability, greenhouse
gases, temperature, and dust distribution, among other things.  While
a great deal of sophisticated and creative work has been done on these
records, very little of that work has leveraged the power of
information theory.  The Shannon entropy rate
\cite{shannon48,Cover:1991:EIT:129837}, for instance, measures the
average rate at which new information---unrelated to anything in the
past---is produced by the system that generated a time series.  If
that rate is very low, the current observation contains a significant
amount of information about the past; conversely, if it is very high,
most of the information in the observation is completely new: i.e.,
the past tells you little or nothing about the future.

This technique can bring out valuable new information from
paleoclimate data records.  Here, we use information theory on the
longest continuous and highest-resolution water-isotope record yet
recovered from Antarctica: the West Antarctica Ice Sheet (WAIS) Divide
core.  We show that the Shannon entropy rate of these data correlates
with accumulation (meters of ice equivalent per year) at the drilling
site, reveals possible signatures of geothermal heating at the base of
the core, and clearly brings out laboratory and data-processing
effects that are difficult to see in the raw data.  These
information-theoretic calculations not only corroborate known facts
and reveal hidden problems with the data, but also suggest new and
sometimes surprising geoscience, and pave the way towards
more-advanced interhemispheric entropy comparisons that could
elucidate some deeper questions about the larger climate system.  The
signatures of Dansgaard-Oeschger events in the information record are
small, for instance, suggesting that these large, abrupt events may
not represent significant changes in the climate system dynamics.

To our knowledge, this paper, and the associated pilot
study~\cite{joshua-ida-2016}, is the first information-theoretic
analysis of an ice-core record\footnote{That study offered preliminary
  evidence that WPE calculations on the WAIS core data were useful in
  identifying data-processing issues and potentially also in
  scientific analysis.}.  Several useful applications of
various entropic measures to time-series data about Earth's {\sl
  current} climate are reviewed in \cite{balasis2013statistical}, and
there is a single published study that used the Shannon entropy rate
to explore different climate-change events captured in Laguna Pallcacocha sedimentary data~\cite{Saco2010}.  By
elucidating how information is created and propagated through the
climate system, information-theoretic studies of ice-core data could
help us identify and understand triggers, amplifiers, sources of
persistence, and globalizers of climate change \cite{Alley03,White14}.

\section*{Materials and Methods}\label{sec:methods}

\subsection*{Ice Core Data Collection and Description}

For the analysis reported here, we used the ratios of $^2${H}$/^1$H
and $^{18}$O$/^{16}$O from the West Antarctic Ice Sheet Divide core
(WDC), abbreviated \dd and \dox, respectively.  The record was
analyzed using a Picarro Inc. cavity ring-down spectroscopy (CRDS)
instrument, coupled to a continuous flow analysis (CFA) system
\cite{cfa-las}.  The data are reported in delta ($\delta$) notation
relative to the baseline Vienna Standard Mean Ocean Water (VSMOW)
and normalized to the Standard Light Antarctic Water (SLAP, \dox $=
-55.5$ \permil, \dd $= -428.0$ \permil) scale.  The $\delta$ values
were determined by $\delta=1000(R_{sample}/R_{VSMOW} - 1)$, where $R$
is the isotopic ratio $^{18}$O$/^{16}$O or D/H (i.e., $^2H/^1H$). The
CRDS-CFA system introduces a noise level of
$\sigma_{noise}=0.55\permil$ for \dd and $\sigma_{noise}=0.09\permil$
for \dox throughout the signal.  
\dox and \dd are proxies for local
temperature and regional atmospheric circulation
resulting from variability in the hydrologic cycle.

Water-isotope traces measured on the Picarro instrument were recorded
at a rate of 1.18 Hz (0.85 s intervals).  Ice samples were moved
through the CRDS-CFA system at a rate of 2.5 cm/min, yielding
millimeter resolution.  The data were then averaged over
non-overlapping 0.5 cm bins.  For each of these data points, an age
was determined using the WDC depth-age scale, providing climate data
from 0--68 \kybp \cite{WDC-isotope-data-paper}.  Annual dating of this
record extends to 31 \kybp\cite{Sigl16}, with the remainder relying on
tie points to the Hulu Cave timescale \cite{Buizert2015}.

\subsection*{Entropy Rate Estimation}

The Shannon entropy rate \cite{shannon48} is typically calculated from
{\sl categorical} data: sequences of symbols, like heads and tails for
a coin-flip experiment.  To calculate it from continuum data like \dd
and \dox, one must first convert those data into symbols.  The typical
approach to this---binning---introduces bias and is fragile in the
face of noise \cite{bollt2001,KSG}.  The permutation entropy of
\cite{bandt2002per} solves that problem using ordinal analysis, which
involves mapping successive elements of that time series to
value-ordered permutations of the same size.  For example, if
successive values of a time series $x_i$ are $(x_1, x_2, x_3) = (6, 1,
4)$ then the ordinal pattern, $\phi(x_1, x_2, x_3)$, of this
three-letter ``word''---formally, a {\sl permutation}---is 231 since $x_2
\leq x_3 \leq x_1$.  The ordinal pattern of the permutation $(x_1,
x_2, x_3) = (60.1, 15.8, 4.0)$ is 321.  By calculating statistics on
the appearance of these permutations in a sliding window across a
signal, one can assess its predictability---that is, how much new
information appears at each time step, on the average, in that segment
of the time series \cite{josh-pre,Pennekamp350017}.

Formally, given a time series $\{x_i\}_{i = 1,\dots,N}$, there is a
set $\mathcal{S}_\ell$ of all $\ell!$ permutations $\pi$ of order
$\ell$.  For each $\pi \in \mathcal{S}_\ell$, one defines the relative
frequency of that permutation occurring in $\{x_i\}_{i = 1,\dots,N}$:
  \begin{equation}
    p(\pi) = \frac{\left|\{i|i \leq N-\ell,\phi(x_{i+1},\dots,x_{i+\ell}) = \pi\}\right|}{N-\ell+1}
  \end{equation}
  where $p(\pi)$ quantifies the probability of an ordinal and
  $|\cdot|$ is set cardinality.  The permutation entropy of order
  $\ell \ge 2$ is:
  \begin{equation}
    \textrm{PE}(\ell) = - \sum_{\pi \in \mathcal{S}_\ell} p(\pi) \log_2 p(\pi)
  \end{equation}
Since $0\le PE(\ell) \le \log_2(\ell!)$ \cite{bandt2002per}, it is
common in the literature to normalize by $\log_2(\ell!)$, producing PE
values that range from 0 to 1.

Note that permutation entropy, as defined above, does not distinguish
between $(x_1, x_2, x_3) = (6, 1, 4)$ and $(x_1, x_2, x_3) = (1000, 1,
4)$ and so it can fail if the observational noise is larger than the
trends in the data but smaller than its large-scale features.  One can
address this issue by introducing a weighting term into the
calculation.  This variant of the technique---weighted permutation
entropy or WPE \cite{fadlallah2013}---is used for all calculations in
this paper, again with a normalization that causes the resulting
values to run from zero (no new information; fully predictable) to 1
(all new information; completely unpredictable).

To calculate WPE as a function of time, one must have evenly sampled
data.  This is a major issue here because the 0.5 cm spacing of the
samples, combined with the nonlinear age-depth relationship of the
core, produce a data series whose temporal spacing increases with
depth\footnote{One could certainly calculate WPE on these data, but
  the timeline of the result would be deformed.}.
In order to calculate WPE from these data, we used linear
interpolation to achieve uniform 1/20th yr spacing.  This simple
approach is not without issues, as linear interpolation introduces
ramps in the signal: repeating patterns in the permutations that can
skew their distribution and thereby lower the WPE.  Note that this
effect will generally worsen with depth because the percentage of
interpolated points in the 1/20th-year spaced versions of the \dd and
\dox traces grows nonlinearly with the depth of the core.  The
specific form of this effect will also depend on the shape of the
climate signal.  The mathematics of information theory currently
offers no way to approach any kind of closed-form derivation of these
complicated effects.  In the face of this, it is important to be
mindful of interpolation-induced effects in WPE calculations.  Among
other things, one should not compare WPE values of a single trace from
an ice core across wide temporal ranges if the data have undergone
depth-dependent interpolation, especially when one is working deep in
the core.  Sampling issues in geoscience data have been the focus of
increased attention in the past few years
\cite{sakellariou2016counting,mccullough2016counting}, and
sophisticated methods have been proposed for working around the
associated problems \cite{boers2017complete,eroglu2016see}.  A careful
study of the relative effects of all of these methods---including the
linear-interpolation approach used here, which is the standard
practice in paleoclimate data analysis---is beyond the scope of this
paper.

Successful use of the WPE method also requires good choices for its
three free parameters: the delay $\tau$ between samples, the word
length $\ell$, and the size $W$ of the sliding window over which the
statistics are calculated for each WPE value.  (In the examples above,
$\ell=3$ and $\tau=1$.)  Very little mathematical guidance is
available for these choices and their effects are not independent.
The $\tau$ parameter controls the spacing of the permutation elements.
For low $\tau$ values, permutations are strongly affected by
high-frequency deviations; for larger $\tau$, those deviations are
filtered out.  The window size $W$ controls the resolution of the
analysis.  The word length $\ell$ must be long enough to allow the
discovery of forbidden ordinals, yet small enough that reasonable
statistics over the ordinals can be gathered in a window of that size.
Choices for the window size and the word length are thus in some
tension, since one generally wants the best possible temporal
resolution.  In the literature, $3 \le \ell \le 6$ is a standard
choice, generally without any formal justification.  In theory, the
permutation entropy should converge to the Shannon entropy rate as
$\ell\to\infty$, but that requires an infinitely long time series~\cite{amigo2012permutation,amigo2005permutation}.  In
practice, the right thing to do is to calculate the {\sl persistent}
permutation entropy by increasing $\ell$ until the large-scale
features of the resulting curve converge. 
That approach was used to choose $\ell = 4$ for the calculations in
this study.  This value represents a good balance between accurate
ordinal statistics and finite-data effects.  That $\ell$ value, in
turn, dictated a minimum window size of 2400 points if one considers
100 counts per ordinal as sufficient \cite{josh-pre}.
This translates to 120 years' worth of ice in the 1/20th-year spaced
WDC traces used here.  

The WAIS Divide ice core is the first such record that is suitable for
this type of information-theoretic analysis.  With a shorter data set,
or one with lower resolution, there would be no way to balance the
tradeoffs outlined in the previous paragraph regarding good choices
for the free parameters of the WPE method.  And there are other
considerations as well: the mathematics of WPE require that each
window samples a weakly stationary system \cite{bandt2002per}. 
This makes the temporal resolution of the sampling even more critical.
The advent of the CFA technology brings ice-core data into the realm
of possibility for this kind of analysis.  It is worth mentioning
that, at the time of this writing, a half-dozen cores are currently
being sampled or resampled with this technology.  All of these records
would be suitable for this type of analysis.

\section*{Results and Discussion}

\begin{figure}[!b]
\begin{center}
\includegraphics[width=1\textwidth]{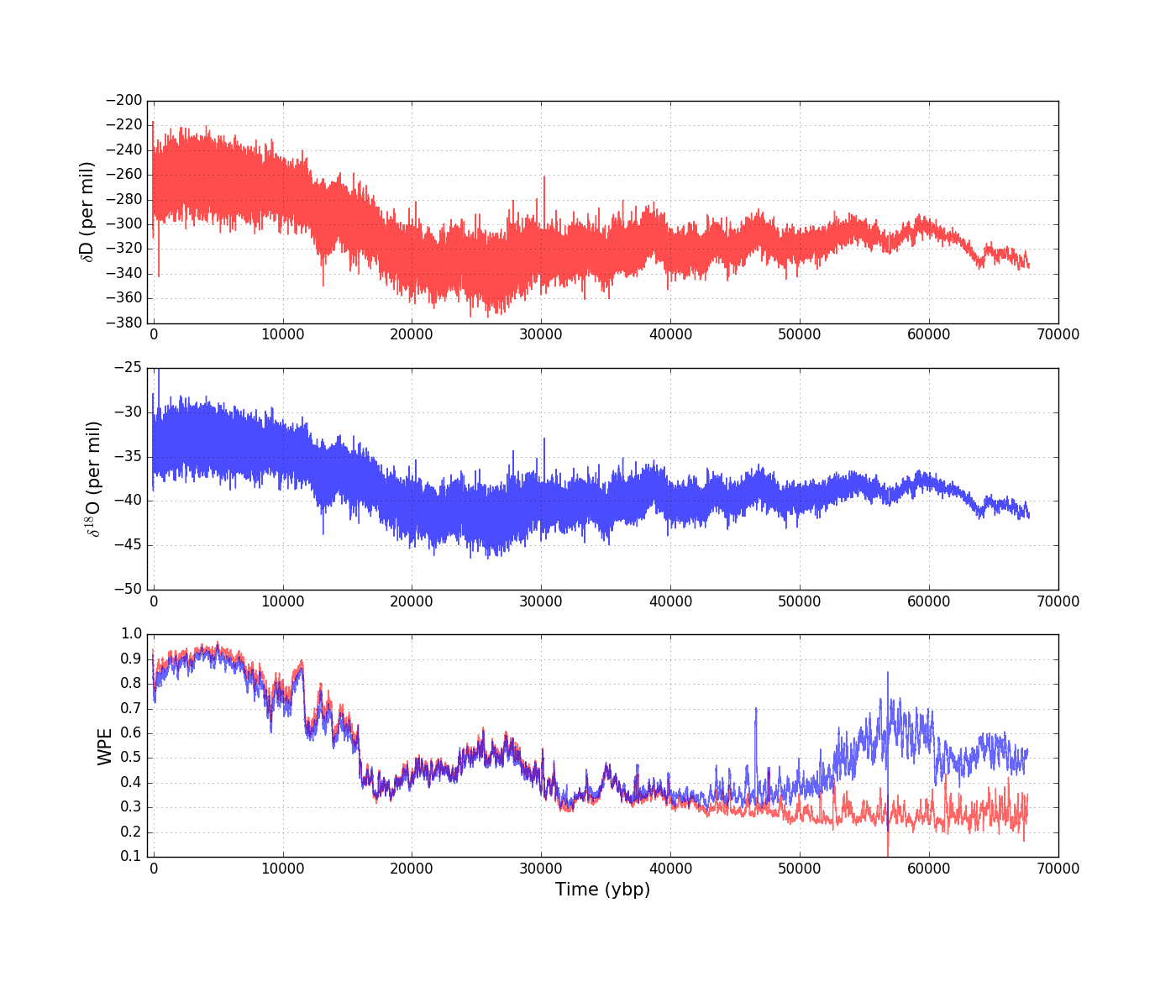}
\caption{{\bf WAIS Divide Core: water isotope values and permutation
  entropy.}  The \dd and \dox data and the weighted permutation entropy
  (WPE) calculated from those data are shown in color (red for \dd and
  blue for \dox). 
The large spikes in both WPE traces near 58 \kybp are due to a
110-year gap in the isotope record.
}
\label{fig:first}
\end{center}
\end{figure}

WPE calculations on the \dd and \dox data from the WDC, shown in
Fig~\ref{fig:first}, reveal how much new information appears, on
average, in a sliding 120-year window leading up to each time point.
A number of features stand out here. The WPE values of both isotopes
are much lower during the glacial period, for instance, than in the
last 5000 years, indicating a stronger dependence of each isotope
value on its previous values.  During the transition from the glacial to the interglacial, both WPE
traces rise sharply at first, beginning around 17 \kybp, then fall
during the Antarctic Cold Reversal (ACR) period from $\approx
14.5-12.9$ \kybp before peaking at the time of the transition from the Younger Dryas
($\approx 12.8-11.5$ \kybp) to the current Holocene period,
coincident with the time of a known spike in accumulation
\cite{GRL:GRL54270}.  This alignment of changes in WPE with known
shifts in the climate system suggests that this technique is
extracting meaningful information from the paleorecord.  As we will
show, there are other features in WPE that correlate with known
climate information---most strongly, accumulation. There are also distinct {\sl differences} between the two WPE traces,
particularly in ice older than 45 \kybp.  Some of these correlations
and disparities, we will argue, may be scientifically meaningful.

\begin{figure}[tb]
\begin{center}
\includegraphics[width=1.0\textwidth]{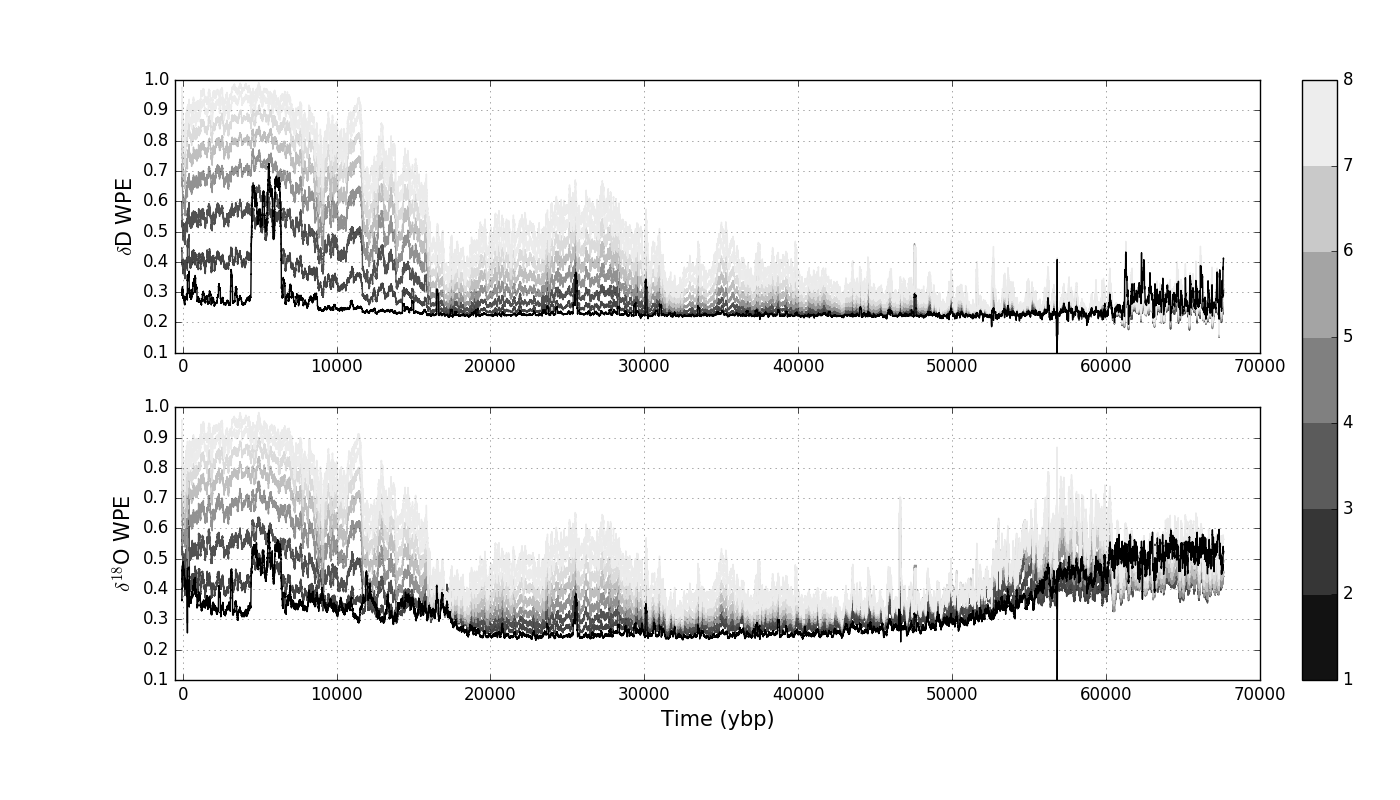}
\caption{{\bf The effects of $\tau$ on WPE.} \dd and \dox WPE
  calculated for different sample spacings $1< \tau < 8$, with the
  $\tau=1$ trace shown in black and larger $\tau$ values in
  successively lighter shades of grey.  (Fig~\ref{fig:first} uses
  $\tau=7$.)  }
\label{fig:tau-sweep}
\end{center}
\end{figure}

The $\tau$ parameter in the WPE formula, which controls the ``stride''
of the calculation, plays a role similar to that of the cutoff
frequency of a low-pass filter.  Fig~\ref{fig:tau-sweep} shows a
series of \dd and \dox WPE calculations with a range of $\tau$ values.
Since the $\tau$ at which a feature disappears is related to the time
scales of the associated effect, one can preferentially focus on---and
distinguish between---long-term effects (e.g., climate) or faster ones
(e.g., weather) simply by tuning $\tau$.  WPE calculations can also
reveal the presence of noise in a signal, as is clear from the large
bump from 4.5-6.5\kybp in the black traces in
Fig~\ref{fig:tau-sweep}.  An older instrument was used to analyze the
ice in this region; closer examination of the data revealed that, as
the WPE results suggested, that instrument introduced significant
noise into the data.
Note that this instrument noise is not visually apparent in the \dd
and \dox measurements in Fig~\ref{fig:first}: that is, WPE is
extracting new information from these data.  Increasing $\tau$
generally raises the WPE curves; this simply reflects decreasing
predictability over the longer time span sampled by each
permutation.  A feature that persists across a range of $\tau$ values,
then, such as the set of bumps between $\approx 9-14$ \kybp in
Fig~\ref{fig:tau-sweep}, indicates an effect in the underlying signal
that spans multiple time scales.  Broad-band {\sl noise} manifests
somewhat differently: as a jumbled set of WPE curves with no clear
trend with $\tau$ (cf., in both \dd and \dox traces below $\approx$ 60
\kybp)\footnote{Signals that are temporally shuffled also produce this
  WPE pattern.}.  All of these patterns---a sharp shift at a
particular $\tau$ value for a band-limited effect, clear features that
persist across a range of $\tau$s for effects that span multiple time
scales, and jumbles of curves for broad-band noise or temporal
shuffling---are recognizable and diagnostic.

\begin{figure}
\includegraphics[width=0.8\textwidth]{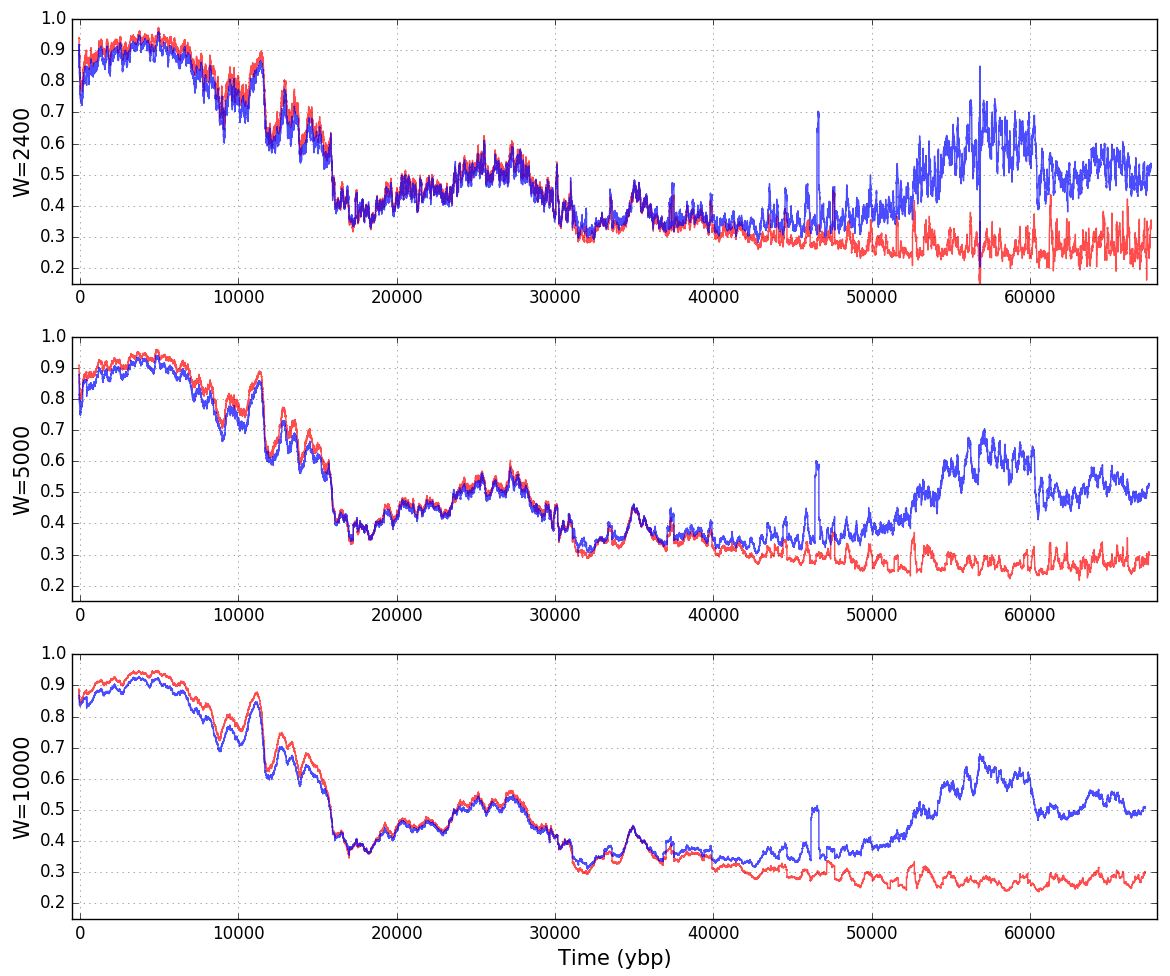}
\caption{{\bf The effect of the window size on WPE.}
  Since WPE aggregates the statistics of the permutations across the calculation window,
  smaller $W$ values increase the variance and larger $W$ values
  smooth out the curves---but without changing their overall features.
  Note that above $W=2400$, this smoothing removes the 57 \kybp spikes
  that were caused by the wide swath of missing isotope data in that
  region.  The bump at $\approx 47$ \kybp is an artifact of a far
  smaller number of outliers in the \dox data; its changing width
  reflects the span of the calculation window in which those points
  play a role.  
}
\label{fig:window-effects}
\end{figure}

WPE also flags other kinds of problems in the data, and in the data
processing.  The spikes around 58 \kybp in Figs~\ref{fig:first} and
\ref{fig:tau-sweep} offer one compelling example.  In this region,
1.107 m (110.1 yr) of ice was missing from the record.  Interpolating
across this gap with a 1/20th year spacing introduced $\approx$ 2387
points, in the form of a linear ramp with positive slope: in other
words, a long series of ``1234'' permutations.  This causes a drop in
WPE as the calculation window passes across this expanse of
interpolated, highly predictable values.  For calculations with $\tau
= 1$ and $W = 2400$, there is a brief period where 99.45\% of the
``data'' in that window has the same permutation, which causes WPE to
fall precipitously, then spike back up as the window starts to move
back onto non-interpolated data.  Larger $\tau$ values---the grey
traces in Fig~\ref{fig:tau-sweep}---mitigate this effect because they
widen both the spacing and the span of the permutation.  Larger window
sizes also mitigate this effect (see \nameref{fig:window-effects}),
but they also decrease the temporal resolution of the WPE analysis.

Outliers in the data that are all but invisible in \dd and \dox traces
also leave clear signatures in WPE, in the form of square waves that
are the width of the calculation window---e.g., at $\approx 48$ \kybp
in the \dd trace in Fig~\ref{fig:tau-sweep}.  A deeper study of the
utility of WPE as an anomaly detection method, complete with
comparisons to standard methods and experimental validation via
targeted re-sampling of a section of the core, appears in a separate
paper \cite{entropy-arxiv}.

Another notable feature in both Figs~\ref{fig:first}
and~\ref{fig:tau-sweep} is the divergence between the \dd and \dox WPE
traces near the base of the ice sheet.  This could be a
data-processing effect due to signal-to-noise ratio, uncertainty in
the age-depth model, and/or interpolation---all of which are more of
an issue this deep in the core, where the signal strengths are lower,
the age-model less certain, and more of the points in the time series
are interpolated.  Since the interpolation and depth-age conversion
processes are identical for the two isotopes, though, they can be
ruled out as possible causes for the divergence in the information
production in \dd and \dox.  The signal-to-noise ratio is also
unlikely to be the cause, though the explanation is more subtle:
diffusion causes the amplitude of the \dd and \dox signals to diminish
with depth.  Since the amount of noise introduced by the measurement
apparatus is constant across the entire core \cite{cfa-las}, the
signal-to-noise {\sl ratio} decreases with depth---possibly
differentially, since \dd is significantly larger than \dox.  However,
the signal-to-noise ratio decreases for \dd and \dox at similar rates
across the whole core: that is, there is no correlated drop in the
signal-to-noise ratio of \dox near the divergence point.

Having ruled out these obvious data and data-processing issues as
causes for the divergence in WPE between \dox and \dd at the deepest
levels of the core, we suspect that this divergence must be related to
something fundamental about \dox and \dd at that depth.  One possible
explanation is the different molecular masses of the two isotopes,
which would cause them to be affected differentially by second-order
thermal effects, e.g., thermal diffusion due to geothermal heat at the
bedrock-ice interface.  This is speculative, of course, and needs
further investigation.  The power of WPE here is its ability to bring
out new knowledge from these traces.  Even though it cannot elucidate
the underlying mechanisms that produced this divergence in information
production between the two isotopes, its ability to bring out that
effect---and suggest hypotheses about its causes---can be useful to
paleoclimate experts.

\begin{figure}[tb]
\begin{center}
\includegraphics[width=1.0\textwidth]{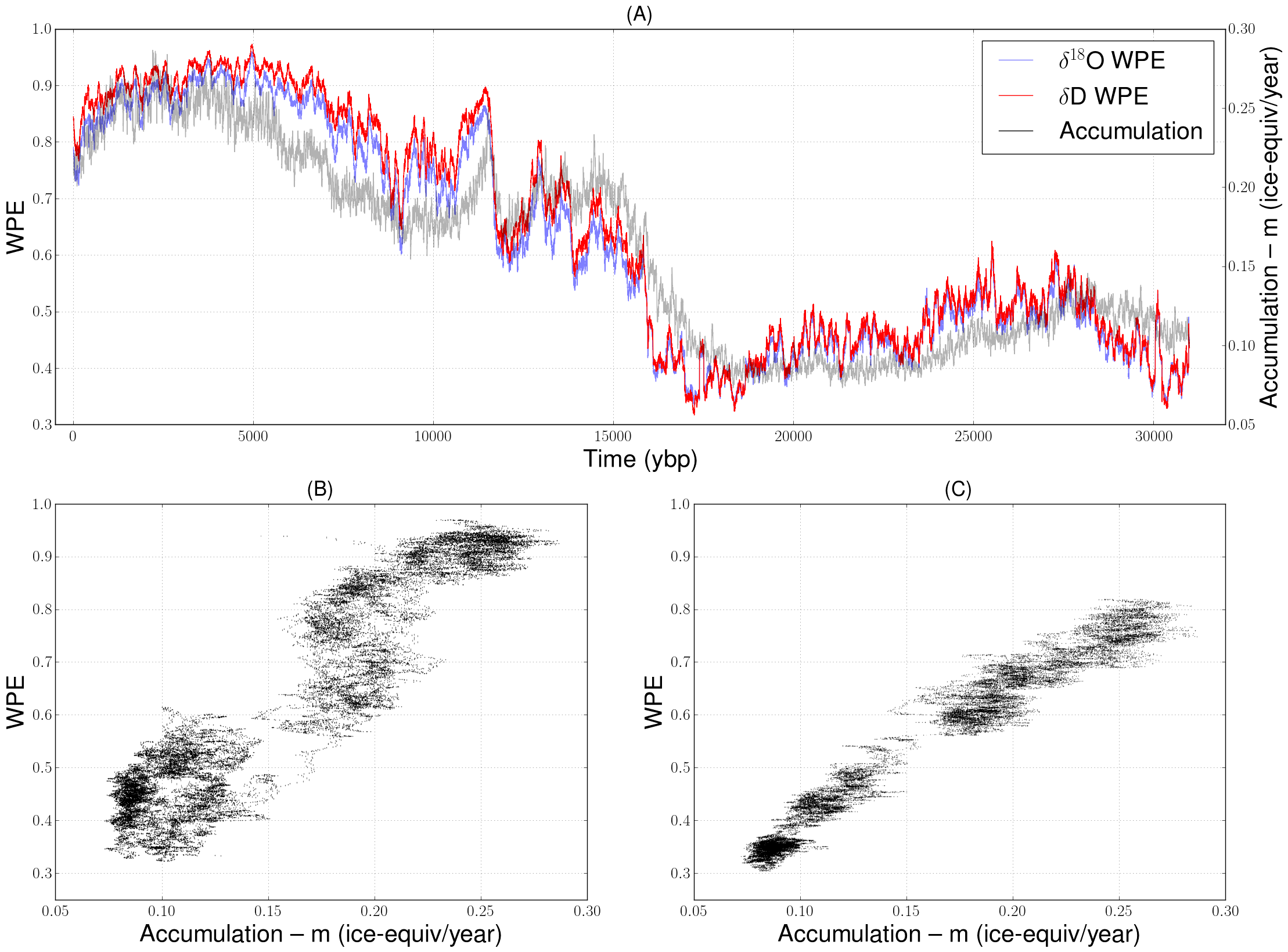}
\caption{{\bf WPE and accumulation.}  (A) time-series traces of \dd
  WPE (red), \dox WPE (blue), and accumulation (grey) at the WAIS
  Divide. 
 Bottom: correlation plots of accumulation and \dd WPE values from (B)
the WDC and (C) a Community Firn Model run~\cite{firnmice}
respectively.  See \nameref{S1_Appendix} for details of this modeling
run (\#2).  Perfect correlation on the bottom two plots would be a
diagonal line.  All signals were averaged over overlapping 25-year
bins to bring out the structure of the plots.  We only show data from
0--31 \kybp because layers cannot be counted below that level, making
it impossible to calculate accumulation values directly.
}
\label{fig:accum}
\end{center}
\end{figure}

Another interesting scientific finding brought out by WPE is the
relationship between WPE and accumulation, which is explored in Fig~\ref{fig:accum}.
Visual examination of part (a) of this figure suggests significant
correspondence between the features: many of the bumps, troughs, and
trends in the three curves occur at the same times in the
record\footnote{The overall pattern of low and high accumulation in
  the glacial and interglacial periods is well known
  \cite{GRL:GRL54270}.}  From first principles, it is not surprising
that WPE tracks accumulation: isotope diffusion intermingles the
information in neighboring layers of the core, which will lower the
WPE.  But there are spatial scales involved in that process, since the
diffusion rate depends on the density of the ice.  And if the annual
layer is thicker, less of the information in that layer will be lost.
In other words, accumulation mitigates diffusion effects, thereby
preserving the information that was laid down in the core.  This means
that the low WPE value during the glacial period may not imply that
the climate was more predictable then; rather, this may simply be due
to lower accumulation.  Diffusion effects may also be the reason why
the curves in Fig~\ref{fig:tau-sweep} cluster tightly in some
regions---e.g., near 17.5 \kybp, where there is a sharp spike in the
diffusion rate at the WDC \cite{JGRF:JGRF20648}.

WPE does not track accumulation perfectly, though; it plateaus earlier
in the Holocene, for instance, and contains some structure during the
glacial period that is not present in the accumulation trace.
The correlation plot in Fig~\ref{fig:accum}(b) explores these
relationships in more detail.  While the WDC results do show a general
trend with accumulation, it is not entirely linear ($R^2=0.927$).  We
conjecture that these deviations from linearity are encodings of
climate signal.  To explore this, we obtained a Community Firn Model
run \cite{firnmice} with the accumulation rate and temperature set to
that measured at the WAIS divide \cite{GRL:GRL54270} and the
water-isotope input fixed throughout the record at a constant annual
amplitude and no variation in the mean.  See \nameref{S1_Appendix} for
the details of this computation.  The results show a more-strongly
linear relationship ($R^2=0.968$) between modeled WPE and
accumulation; see Fig~\ref{fig:accum}(c).  That is, WPE and
accumulation are very tightly correlated in a model run that includes
no climate variability.  The obvious deviation from linearity in
Fig~\ref{fig:accum}(b) around 15--16 \kybp occurs during the transition
from the glacial to interglacial periods, where many climatic
variables are known to have changed \cite{WDC-isotope-data-paper}.  In
the CFM, we did not consider the effect of climate changes on the
water isotope signal, so it is not surprising that this deviation is
not present in Fig~\ref{fig:accum}(c).  Indeed, this further confirms
the underlying linear relationship that is suggested by the
climatology, thereby adding weight to the conjecture that the
deviations from linearity may be encodings of climate
signal\footnote{It is worth noting that rudimentary statistics---e.g.,
  rolling-window variance calculations---do not track accumulation and
  do not reveal any divergence between the isotopes near the base of
  the core \cite{entropy-arxiv}.}

\begin{figure}[tb]
\begin{center}
\includegraphics[width=\textwidth]{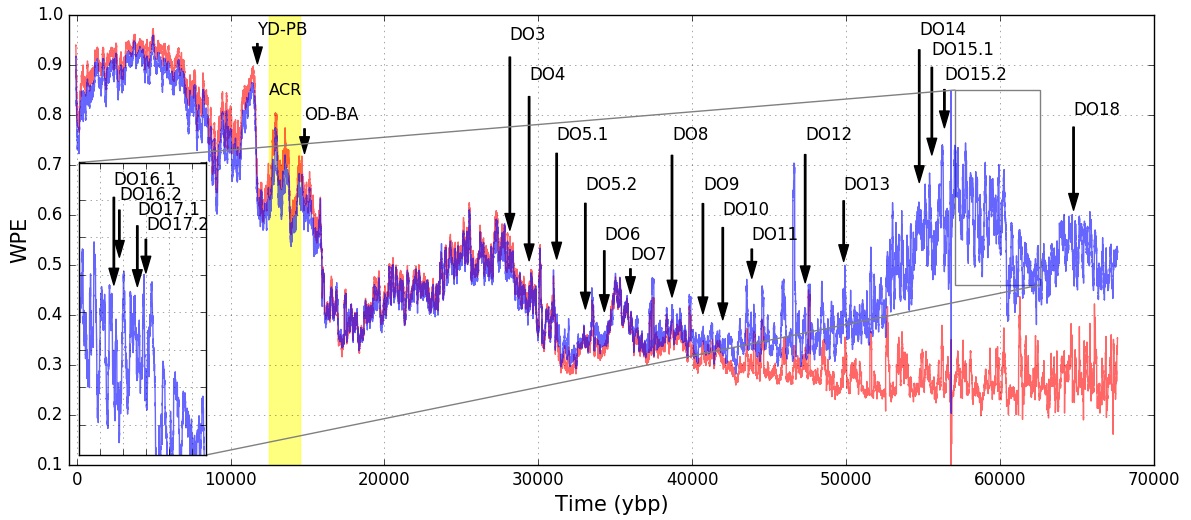}
\caption{{\bf WPE with Dansgaard-Oeschger events labeled for
    comparison.} WPE of \dd (red) and \dox
  (blue) with Dansgaard-Oeschger, Antarctic Cold Reversal, and Younger
  \& Older Dryas events shown. 
}
\label{fig:WPEs-DO-AIM}
\end{center}
\end{figure}

Another interesting property of the WPE traces is what is {\sl not}
there: specifically, there is no systematic correspondence between
features in the WDC WPE traces and Antarctic Isotope Maxima (AIM) \cite{WAIS15} or Dansgaard-Oeschger
(DO) \cite{DO} events; see Fig~\ref{fig:WPEs-DO-AIM}.  There {\sl is}
a clear peak in WPE at the time of the Younger Dryas event, but that
is probably due to the accumulation effects described above.
Spectral analysis of the WPE traces (see Table~\ref{tab:spectral}) shows that while millennial frequencies
persist throughout the records, they are not 99\% statistically
significant.  That is, we do not see concrete evidence of any
persistent frequencies that might correspond to a repeating trigger
mechanism of DO and AIM events.
Rather, the WPE analysis suggests that while these events
substantially changed the temperature signatures in cores from some
regions, they may not have represented substantial changes in the
overall Earth climate dynamics.

Absence of evidence, of course, is not evidence of absence.  This is a
particularly thorny point when one is using nonlinear statistics on
sparse data.  In the literature on WPE, the important issue of
significance---whether or not a given feature in a WPE trace (e.g.,
jump, spike, valley) indicates some sort of substantive change in the
underlying information mechanics of the system---is almost never
addressed.  A very recent paper \cite{Bandt2017} offers some
preliminary solutions to this problem, including a variant of
permutation entropy that measures how far the signal is from white
noise.  
However, this is far from a general solution; with only one time
series available, traditional significance tests from statistics are
inapplicable, and without a reliable generative model for the climate,
one cannot use surrogate methods for significance testing
\cite{lancaster2018surrogate}.  While there are several pseudo
significance tests, such as feature persistence over ranges of
parameters, the associated theories are undeveloped.  Traditional
methods like randomized bootstrapping \cite{bootstrapping} may
eventually be useful here, but the associated mathematics has not yet
been extended to information theory.  Until these shortcomings have
been addressed, interpreting small-scale fluctuations---such as those
near the times of DO events in Fig~\ref{fig:WPEs-DO-AIM}---should
either be avoided altogether or done with careful consideration and
persistence testing over a wide range of values for the free
parameters of the algorithm.

\section*{Conclusion}
 
The central claim of this paper is that the climate information
captured in paleorecords can be better understood with the aid of
information theory.  As evidence for this claim, and of the traction
that it can offer on paleoclimatology problems, we demonstrated that
estimates of the Shannon entropy rate of the water-isotope data from
the WAIS Divide Core, calculated using weighted permutation entropy
(WPE), can bring out valuable new information from this record.  We
found that WPE correlates with accumulation,
reveals possible signatures of geothermal heating at the base of the
core, and clearly brings out laboratory and data-processing effects
that are difficult to see in the raw data.  WPE also contains features
that do not correspond to well-known climate phenomena (e.g., DO and
AIM events), nor to features in the accumulation record.  We suspect
that these are encodings of climate signal, but the task of separating
out that information from the accumulation/diffusion effects is a real
challenge because of the complexity of the mathematics of WPE. 
While information-theoretic measures are powerful, they require
careful handling and high-resolution, well-dated data.  Data issues
and pre-processing steps that affect the timeline can skew their
results, as discussed at length in two recent papers
\cite{mccullough2016counting,sakellariou2016counting}.  Moreover,
the associated algorithms have a number of free parameters that must be chosen
properly.  (This is true of any other data-analysis method, of course,
though that is not widely appreciated in many scientific fields.) 
The WAIS Divide ice core is the first ice-core record that is suitable
for these types of analyses.  As more high-resolution records become
available, and the mathematics is developed, information theory will
likely become a common forensic tool in climate science.  

\section*{Acknowledgments}

This material is based upon work sponsored by the National Science
Foundation (Grant No. 1245947,1807478). For ice core data, this work
was supported by US National Science Foundation (NSF) grants 0537593,
0537661, 0537930, 0539232, 1043092, 1043167, 1043518 and
1142166. Field and logistical activities were managed by the WAIS
Divide Science Coordination Office at the Desert Research Institute,
USA, and the University of New Hampshire, USA (NSF grants 0230396,
0440817, 0944266 and 0944348). The NSF Division of Polar Programs
funded the Ice Drilling Program Office (IDPO), the Ice Drilling Design
and Operations (IDDO) group, the National Ice Core Laboratory (NICL),
the Antarctic Support Contractor, and the 109th New York Air National
Guard. Water isotope measurements were performed at the Stable Isotope
Lab (SIL) at the Institute of Arctic and Alpine Research (INSTAAR),
University of Colorado. Any opinions, findings, and conclusions or
recommendations expressed in this material are those of the author(s)
and do not necessarily reflect the views of the NSF.  JG was supported
by an Omidyar Fellowship by the Santa Fe Institute. The authors would
also like to thank Jakob Runge, Nihat Ay, Holger Kantz and Stephan
Bialonski for valuable discussion, as well as Bruce Vaughn and Valerie
Morris for their efforts in designing the laser-based, continuous flow
measurement system for water isotopes.  Valerie was also instrumental
in the data processing.  We would also like to acknowledge the
reviewers for their unusually thorough and thoughtful feedback on this
work, which significantly improved this paper.


\section*{Supporting information}
\beginsupplement

\begin{table}[tb]
   \begin{center}
   \begin{tabular}{|c|c|c|c|c|c|}
   \hline epoch & $\tau$=1 &$\tau$=3 &$\tau$=5 &$\tau$=7 &$\tau$=9 \\
   \hline 
   glacial (20-30 \kybp) & 655 & 546 & 546 & 546 & 546 \\
   transition (10-20 \kybp) & 728 & 273 & 728 & 771 & 1092, 655 \\
   holocene (0-10\kybp ) & 1191, 728 & 364 & 1008, 596 & 873 & 1008, 624 \\ \hline 
   \end{tabular}
   \caption{{\bf Spectral analysis.} The 99\% significant peaks in the
centennial and  millenial-scale frequency range (350-4500 years).
These results were calculated from the \dd WPE trace in
Figure~\ref{fig:first} using the MTM kspectra package
\cite{kspectra-climate} with standard parameter values (three tapers,
a resolution of 2.0, and a red-noise null model) and a sampling interval
of 0.05 yr to match the timescale of the data.
}
 \label{tab:spectral}
   \end{center}
\end{table}

\paragraph*{S1 Appendix}
\label{S1_Appendix}
{\bf Community Firn Model.} The Community Firn Model (CFM)
\cite{firnmice} was used to investigate the effects of firn processes
on water isotope WPE.  In the CFM, individual packets of snow/firn/ice
are tracked downward over time.  At each time step, a new packet is
added on top, and the oldest packet is removed from the bottom of the
stack.  Each packet is compressed at each time step based on its
overburden load, temperature, and any other tracked parameters in the
model physics.  Temperature is also calculated at each step using
thermal parameters appropriate for the current density-depth
structure.

A synthetic input isotope signal spanning 30 \kybp was created based
on a cosine wave, with an amplitude ($a$) of 2 \permil,
time step ($\Delta t$) of 1/12$^{th}$ yr, and a mean value ($\mu$) of -28
\permil: 
$$\Delta_{\cos} = a \cos(2\pi t) + \mu$$
Red noise was added to $\Delta_{\cos}$ to produce the synthetic isotope
signal $\Delta_{syn}$ for the CFM run:
$$\Delta_{syn}(i+1) = k  \Delta_{\cos}(i) + w(i); $$
where $k=0.7$
and $w$ 
is white noise with a standard deviation of 20\% of the annual
amplitude of the cosine wave. As observed at WAIS Divide, the model
depth in the ice sheet at 30 kyr was set to 2816.435 m, and the model
depth at the base of the ice sheet was set to 3405 m.

We analyzed two different CFM scenarios, all using the $\Delta_{syn}$
signal described above:
\begin{enumerate}
\item firn density of 400 kg/m$^3$, estimated temperature at WAIS
  Divide \cite{cuffey16}, thermal diffusion ON, estimated accumulation
  at WAIS Divide \cite{GRL:GRL54270}, and isotope diffusion OFF
\item firn density of 400 kg/m$^3$, estimated temperature at WAIS
  Divide \cite{cuffey16}, thermal diffusion ON, estimated accumulation
  at WAIS Divide \cite{GRL:GRL54270}, and isotope diffusion ON
\end{enumerate}

\begin{figure}[h]
\begin{center}
\includegraphics[width=0.9\textwidth]{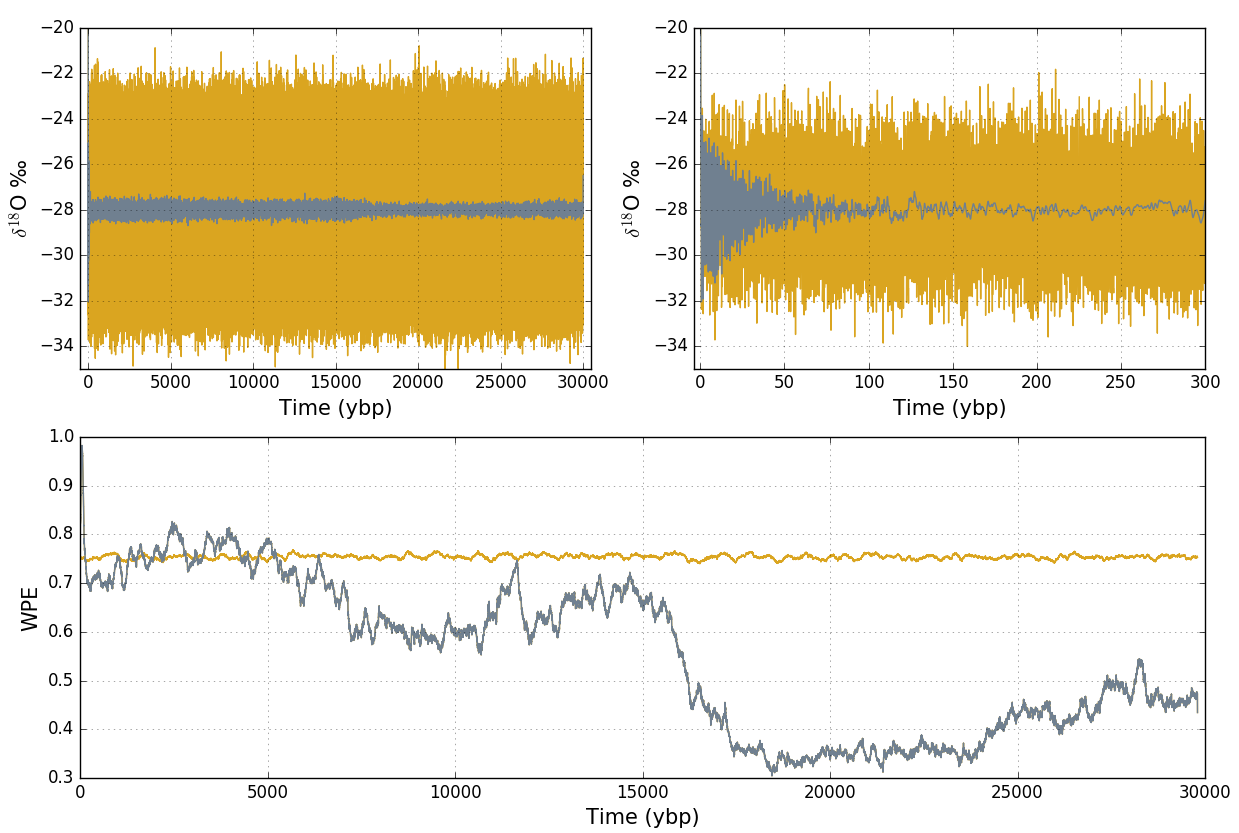}
\caption{{\bf Community Firn Model Results.} (a) Synthetic isotope output data
for Community Firn Model (CFM) experiment 1 (gold) and experiment 2
(grey) (see S1 Appendix).  The amplitude of the data in experiment 2 is
reduced due to the inclusion of isotopic diffusion in the firn.  (b)
As in (a), for the most recent 300 years of data.  The grey line shows
the effects of firn diffusion, which increasingly reduces the signal
amplitude until the bubble close-off depth. (c) WPE of CFM output data
for experiment 1 (gold) and experiment 2 (grey). The WPE in experiment
2 is very similar to the WDC accumulation rate, due to the combined
effects of accumulation and isotopic diffusion in the CFM.  See
\nameref{S1_Appendix} for more details about these simulations.}
\label{fig:cfm}
\end{center}
\end{figure}

The output for these experiments can be seen in the top two panels of
Fig~\ref{fig:cfm}.  The second iteration of the CFM model run (shown
in grey in Fig~\ref{fig:cfm}) is used in Fig~\ref{fig:accum}.  The
bottom panel of Fig~\ref{fig:cfm} reiterates the strong correlation
that is described in the Results section between accumulation and WPE
when isotopic diffusion is present and the complete lack of
correlation when isotopic diffusion is not present.

\nolinenumbers

\clearpage
\bibliography{master-refs}

\end{document}